\documentclass[fleqn,colordvi, 11pt] {article}
\usepackage{amsthm}
\usepackage{graphicx, amsmath, amssymb, citesort}
\usepackage{amsfonts}

\begin{document}

\title {\Large Reducing Tile Complexity for Self-Assembly Through Temperature Programming\thanks{A preliminary form of this paper
will appear
  in \textit{Proceedings of the 17th Annual ACM-SIAM Symposium on Discrete
  Algorithms}, 2006.}}

\author {
        Ming-Yang Kao\thanks{Department of Electrical Engineering and Computer Science, Northwestern University,
Evanston, IL 60208. Email: {\tt kao@cs.northwestern.edu}.  The
research of this author was supported in part by NSF grant
EIA-0112934.}
    \and
        Robert Schweller\thanks{Department of Electrical Engineering and Computer Science, Northwestern University,
Evanston, IL 60208. Email: {\tt schwellerr@cs.northwestern.edu}.
The research of this author was supported in part by NSF grant
EIA-0112934.}
 }

\date{\today}

\maketitle

\newtheorem{theorem}{Theorem}
\newtheorem*{thmmain}{Theorem 1}
\newtheorem{definition}{Definition}
\newtheorem{conjecture}{Conjecture}
\newtheorem{corollary}{Corollary}
\newtheorem{lemma}{Lemma}

\newcommand{\see}[1]{{\scriptsize #1}}
\newcommand{\spaceA}{\mbox{\hspace{0.1in}}}
\newcommand{\spaceB}{\mbox{\hspace{0.1in}}}
\newcommand{\spaceC}{\mbox{\hspace{0.4in}}}
\newcommand{\spaceD}{\mbox{\hspace{0.3in}}}
\newcommand{\spaceE}{\mbox{\hspace{1.25in}}}
\newcommand{\spaceF}{\mbox{\hspace{1.55in}}}

\begin{abstract}
We consider the tile self-assembly model and how tile complexity can
be eliminated by permitting the temperature of the self-assembly
system to be adjusted throughout the assembly process.  To do this,
we propose novel techniques for designing tile sets that permit an
arbitrary length $m$ binary number to be encoded into a sequence of
$O(m)$ temperature changes such that the tile set uniquely assembles
a supertile that precisely encodes the corresponding binary number.
As an application, we show how this provides a general tile set of
size $O(1)$ that is capable of uniquely assembling essentially any
$n\times n$ square, where the assembled square is determined by a
temperature sequence of length $O(\log n)$ that encodes a binary
description of $n$.  This yields an important decrease in tile
complexity from the required $\Omega(\frac{\log n}{\log\log n})$ for
almost all $n$ when the temperature of the system is fixed.  We
further show that for almost all $n$, no tile system can
simultaneously achieve both $o(\log n)$ temperature complexity and
$o(\frac{\log n}{\log\log n})$ tile complexity, showing that both
versions of an optimal square building scheme have been discovered.
This work suggests that temperature change can constitute a natural,
dynamic method for providing input to self-assembly systems that is
potentially superior to the current technique of designing large
tile sets with specific inputs hardwired into the tileset.
\end{abstract}

\section{Introduction}
Self-assembly is the ubiquitous process by which objects
autonomously assemble into complexes.  This phenomenon is common in
nature and yet is poorly understood from a mathematical perspective.
It is believed that self-assembly technology will ultimately permit
the precise fabrication of complex nanostructures. Of particular
interest is DNA self-assembly. Double and triple crossover DNA
molecules have been designed that can act as four-sided building
blocks for DNA self-assembly~\cite{Fu:1993:DDC,LaBean:2000:CAL}.
Experimental work has been done to show the effectiveness of using
these building blocks to assemble DNA crystals and perform DNA
computation~\cite{Lagoudakis:1999:SAS,Reif:1997:LPB,Winfree:1998:DSA,Winfree:1996:UCV}.
With these building blocks (tiles) in mind, researchers have
considered the power of the \emph{tile} self-assembly model.

The tile assembly model extends the theory of Wang tilings of the
plane~\cite{Wang:1961:PTP} by adding a natural mechanism for growth.
Informally, the model consists of a set of four sided Wang tiles
whose sides are each associated with a type of glue.  The bonding
strength between any two glues is determined by a \emph{glue
function}.  A special tile in the tile set is denoted as the
\emph{seed} tile.  Assembly takes place by starting with the seed
tile and attaching copies of tiles from the tile set one by one to
the growing seed whenever the total strength of attraction from the
glue function meets or exceeds a fixed parameter called the
\emph{temperature}.

In this paper we consider the power of the \textit{multiple
temperature} model proposed by Aggarwal et
al.~\cite{Aggarwal:2005:CGM,Aggarwal:2004:CGM}.  The multiple
temperature model generalizes the tile self-assembly model by
permitting the temperature of the system to shift up and down.
Aggarwal et al. showed that by raising the temperature just once
during the assembly process certain shapes could be uniquely
assembled using fewer distinct tiles than what is possible for a
model with a fixed temperature.

We extend these results by showing how tile sets can be designed so
that they assemble supertiles that encode bits of information
specified by a sequence of increases and decreases in temperature.
We provide a general tile set that is capable of uniquely
assembling, for any arbitrary length $m$ binary number, a
corresponding supertile that encodes that number, where the encoded
binary string is specified by a length $O(m)$ sequence of
temperature changes. We then extend this construction by including a
constant size set of generic square building tiles that, after a
final drop in temperature, read the encoded binary number and
uniquely assemble the $n\times n$ square specified by the binary
string. We thus provide a constant size set of tiles that are singly
capable of uniquely assembling any $n\times n$ square, where the
square assembled is determined by a length $O(\log n)$ temperature
sequence which encodes a description of $n$.  This construction
stands in sharp contrast to what is possible for a fixed temperature
system. For almost all $n$, Kolmogorov complexity dictates a lower
bound of $\Omega(\frac{\log n}{\log\log n})$ distinct tiles required
to uniquely assembly an $n\times n$ square for almost all
$n$~\cite{Rothemund:2000:PSC}.

While our focus is on the assembly of $n\times n$ squares, we
believe the technique of temperature programming should be useful
for a number of different self-assembly applications.  Interesting
tile self-assembly systems can usually be broken up into two parts:
An \emph{input} set of tiles of arbitrary size, and an
\emph{execution} set of tiles of constant size.  For example,
Rothemund et al. \cite{Rothemund:2000:PSC} assembled squares using a
set of $O(\log n)$ input tiles that uniquely assemble a $\log n$
digit binary number, combined with a constant size set of execution
tiles that read the produced binary number and uniquely assemble the
corresponding size square. Similarly, tile sets designed for DNA
computation require one group of input tiles of arbitrary length
designed to encode a given input, and another set of constant size
designed for the computation of some procedure such as exclusive-or,
evaluation of 3-SAT,
etc~\cite{Lagoudakis:1999:SAS,Reif:1997:LPB,Winfree:1998:DSA,Winfree:1996:UCV}.
This work initializes the study of exchanging sets of tiles
representing an input for a sequence of temperature changes.

In this paper, we show how to encode the input information for the
assembly of $n\times n$ squares into a sequence of $O(\log n)$
temperature changes, rather than explicitly encoding the input into
the tile system.  This drops the tile complexity of the system to
$O(1)$ and provides a generic tile set that can be programmed via
temperature changes to assemble any $n\times n$ square.  We believe
this natural, dynamic method of providing input to a tile system is
potentially more practical than explicitly creating a large,
hardwired input set of new tiles or DNA molecules for each distinct
input $n$.  Further, we suspect that any tile set that follows the
basic input, execution framework can be adjusted slightly so as to
take input from a short temperature sequence rather than a set of
input tiles, similar to our adaptation of square building tiles.

\vspace{0.1in} $\textbf{Paper Layout:}$ In Section~\ref{sec:basics}
we define the self-assembly model and state the main result of the
paper. In Section~\ref{section:ptstc} we introduce a method of
encoding bits of information into self-assembled shapes by shifting
temperature, which we apply to assemble arbitrary length binary
strings using a general tile set of size $O(1)$.  In
Section~\ref{section:bast} we extend the construction for building
binary strings to one for building arbitrary $n\times n$ squares in
$O(1)$ tiles and $O(\log n)$ temperature changes.  In
Section~\ref{section:ar} we discuss a modification for making our
constructions more robust. In Section~\ref{section:future} we
conclude with a discussion of future research directions.

\section{Basics and Main Result}\label{sec:basics}

\subsection{Definitions}
To describe the tile self-assembly model, we make the following
definitions.  A tile $t$ in the model is a four sided Wang tile
denoted by the ordered quadruple $(\textrm{north}(t),
\textrm{east}(t), \textrm{south}(t), \textrm{west}(t))$. The entries
of the quadruples are glue types taken from an alphabet $\Sigma$
representing the north, east, south, and west edges of the Wang
tile, respectively. A \textit{tile system} is an ordered quadruple
$\langle T, s, G, \tau \rangle$ where $T$ is a set of tiles called
the $\textit{tileset}$ of the system, $\tau$ is a positive integer
called the \emph{temperature} of the system, $s \in T$ is a single
tile called the $\textit{seed}$ tile, and $G$ is a function from
${\Sigma}^2$ to $ \{ 0,1,\ldots ,\tau \} $ called the \emph{glue
function} of the system.  It is assumed that $G(x,y) = G(y,x)$, and
there exists a $\tt{null}$ in $\Sigma$ such that $\forall x \in
\Sigma$, $G(\tt{null},x) = 0$.  In this paper we assume the glue
function is such that $G(x,y) = 0$ when $x\neq y$ and denote
$G(x,x)$ by $G(x)$ (See~\cite{Aggarwal:2005:CGM,Aggarwal:2004:CGM}
for the effect of removing this restriction).  $|T|$ is referred to
as the \emph{tile complexity} of the system.

Define a $\textit{configuration}$ to be a mapping from
${\mathbb{Z}}^{2}$ to $T$ $\bigcup$ $\{\tt{empty}\}$, where
$\tt{empty}$ is a special tile that has the $\tt{null}$ glue on
each of its four edges.  The \emph{shape} of a configuration is
defined as the set of positions $(i,j)$ that do not map to the
empty tile.  For a configuration~$C$, a tile $t \in T$ is said to
be \textit{attachable} at the position $(i,j)$ if $C(i,j) =
\tt{empty}$ and $ G(\textrm{north}(t) , \textrm{south}(C(i,j+1)))+
G(\textrm{east}(t) , \textrm{west}(C(i+1, j)))+
G(\textrm{south}(t) , \textrm{north}(C(i, j-1)))+
G(\textrm{west}(t) , \textrm{east}(C(i-1, j))) \geq \tau$.  For
configurations $C$ and $C'$ such that $C(x,y) = \tt{empty}$,
$C'(i,j) = C(i,j)$ for all $(i,j) \neq (x,y)$, and $C'(x,y) = t$
for some $t\in T$, define the act of \textit{attaching} tile $t$
to $C$ at position $(x,y)$ as the transformation from
configuration $C$ to $C'$.

Define the \textit{adjacency graph} of a configuration $C$ as
follows. Let the set of vertices be the set of coordinates $(i,j)$
such that $C(i,j)$ is not empty.  Let there be an edge between
vertices $(x_1, y_1)$ and $(x_2, y_2)$ iff $|x_1 - x_2| + |y_1 -
y_2| = 1$. We refer to a configuration whose adjacency graph is
finite and connected as a $\textit{supertile}$.  For a supertile
$S$, denote the number of non-empty positions (tiles) in the
supertile by $\textrm{size}(S)$.  We also note that each tile $t
\in T$ can be thought of as denoting the unique supertile that
maps position $(0,0)$ to $t$ and all other positions to
$\tt{empty}$. Throughout this paper we will informally refer to
tiles as being supertiles.

A $\textit{cut}$ of a supertile is a cut of the adjacency graph of
the supertile.  In addition, for each edge $e_i$ in a cut the
$\textit{edge strength}$ $s_i$ of $e_i$ is the glue strength (from
the glue function) of the glues on the abutting edges of the
adjacent tiles corresponding to $e_i$.  The $\textit{cut
strength}$ of a cut $c$ is $\sum s_i$ for each edge $e_i$ in the
cut.  A supertile is said to be \emph{stable} at temperature
$\tau$ if there exists no cut of the supertile with strength less
than $\tau$.

\subsection{The Assembly Process}\label{subsec:standardmodel}
Assembly takes place by \textit{growing} a supertile starting with
tile $s$ at position $(0,0)$.  Any $t \in T$ that is attachable at
some position $(i,j)$ may attach and thus increase the size of the
supertile.  For a given tile system, any supertile that can be
obtained by starting with the seed and attaching arbitrary
attachable tiles is said to be \emph{produced}. If this process
comes to a point at which no tiles in $T$ can be added, the
resultant supertile is said to be \textit{terminally} produced. For
a given shape $\Upsilon$, a tile system $\Gamma$ \textit{uniquely
produces} shape $\Upsilon$ if there exists a terminal supertile $S$
of shape $\Upsilon$ such that every supertile derived from the seed
can be grown into $S$. The $\textit{tile complexity}$ of a shape
$\Upsilon$ is the minimum tile set size required to uniquely
assemble $\Upsilon$.

\subsection{The Multiple Temperature Model}\label{s:FTM}
In the $\textit{multiple temperature}$ model, the integer
temperature $\tau$ in the tile system description is replaced with
a sequence of integers $\{ \tau_i \}_{i=1}^{k}$ called the
$\textit{temperature sequence}$ of the system.  The number of
temperatures $k$ in the sequence is called the \emph{temperature
complexity} of the system.  The size of the largest temperature in
the sequence is called the \emph{temperature range} of the system.
In this paper, we require the temperature range to be bounded by a
constant.

In a system with $k$ temperatures, assembly takes place in $k$
phases. In the first phase, assembly takes place as in the
standard model under temperature $\tau_1$.  Phase 1 continues
until no tiles can be added.  In phase two, tiles can be added or
removed under $\tau_2$. Specifically, at any point during phase 2,
if there exists a cut of the resultant supertile with cut strength
less than $\tau_2$, the portion of the supertile occurring on the
side of the cut not containing the seed tile may be removed. Also,
at any point in the second phase any tile in $T$ may be added to
the supertile if the tile is attachable at a given position under
temperature $\tau_2$. The second phase of this assembly continues
until no tiles can be added or removed. We then go to phase 3 in
which tiles may be added and removed under temperature $\tau_3$.
The process is continued up through $\tau_k$. For any given choice
of additions and removals such that each of the $k$ phases
finishes, the tile system \textit{terminally} produces the
corresponding shape assembled after phase $k$. If the $k$ phases
always finish regardless of the choice of additions and removals,
and the terminally produced supertile $R$ is unique, the tile
system \textit{uniquely assembles} the shape of $R$.

\subsection{Main Result}
The main result of this paper is the following theorem which shows
that there exists a single tile set of constant size that can be
efficiently programmed by a series of temperature changes to
uniquely assembly essentially any $n\times n$ square.

\begin{theorem}\label{theorem:square}  There exists a tile set $T$ of size $O(1)$ such
that for any $n \geq 22$ there exists a temperature sequence
$\{\tau_i\}_{i=1}^{k}$ of length $k=O(\log n)$ such that $T$
uniquely assembles an $n\times n$ square with temperature sequence
$\{\tau_i\}_{i=1}^{k}$.
\end{theorem}

This theorem is proven in~Section~\ref{section:bast}.  We also show
in Section~\ref{section:bast} that for almost all $n$, a tile system
that assembles an $n\times n$ square cannot simultaneously achieve
$o(\frac{\log n}{\log\log n})$ tile complexity and $o(\log n)$
temperature complexity.  This shows that there does not exist a
smooth tradeoff between tile complexity and temperature complexity,
and that our result constitutes the optimal achievable scheme for
any system that hopes to drop tile complexity below the
$O(\frac{\log n}{\log\log n})$ bound achieved
in~\cite{Adleman:2001:RTP}.

To design the tile set for Theorem~\ref{theorem:square}, we first
introduce in Section~\ref{section:ptstc} a technique for encoding
bits of information into assembled supertiles via temperature
shifts, thus permitting the assembly of supertiles that encode
arbitrary binary strings.  We then extend this construction in
Section~\ref{section:bast} by adding a constant size set of square
building tiles that read a binary string encoded into an input
supertile and assemble a corresponding $n\times n$ square.  We
thus show that it is possible to assemble any $n\times n$ square
in $O(1)$ tile complexity and $O(\log n)$ temperature complexity.
This is in contrast to the tile complexity lower bound of
$\Omega(\frac{\log n}{\log\log n})$ that holds for almost all $n$
for any system that only has a constant number of temperature
shifts~\cite{Rothemund:2000:PSC,Aggarwal:2004:CGM}.

\section{Programming a Tile Set with Temperature Change}
\label{section:ptstc}

In this section we introduce the \emph{bit-flip gadget} which
permits a bit of information to be encoded into a produced
supertile by shifting the temperature of the self-assembly system.
We then show how these gadgets can be used to encode arbitrary
length $m$ binary strings with a fixed size tile set and $O(m)$
temperature changes.

\subsection{Bit-Flipping}
\begin{figure*}[t]
\label{figure:GadgetTiles}
\begin{center}
\includegraphics*[scale=.50]{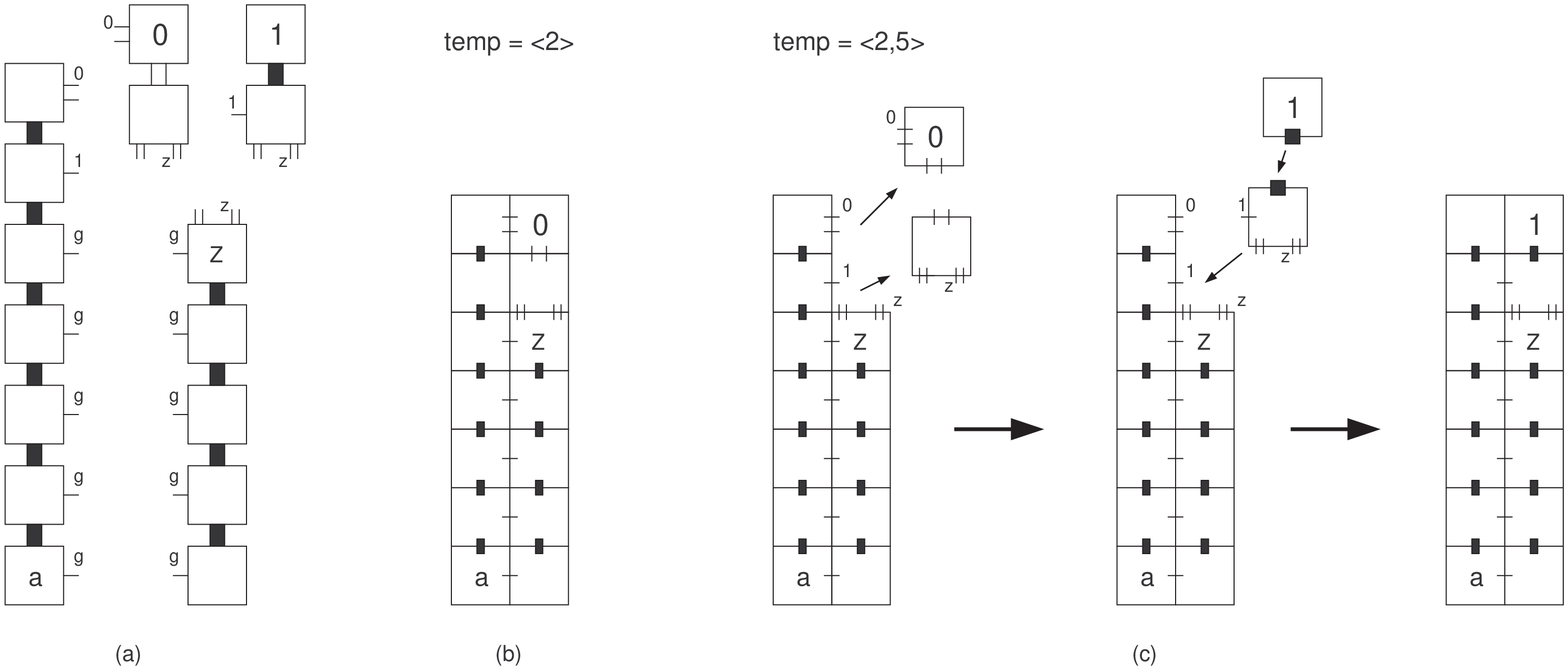}
\caption{(a)  Tiles that implement the bit flip gadget.  The number
of lines on the side of a tile represents the strength of the
corresponding glue.  The dark black line denotes a strength 5 glue.
(b) At temperature 2, with tile $a$ as the seed tile, the bit flip
gadget uniquely assembles a supertile with the 0 tile in the top
right corner.  (c) By raising the temperature to 5, the 0 tile
breaks off and is replaced by the 1 tile.}
\label{figure:GadgetTiles}
\end{center}
\end{figure*}

The basic tool used in our constructions in this paper is a set of
tiles we collectively refer to as a \emph{bit-flip gadget}.  The
tiles for the basic gadget are depicted in
Figure~\ref{figure:GadgetTiles}. The idea is that the gadget
uniquely assembles a shape that encodes a binary 0 for an initial
temperature, then flips the bit and encodes a binary 1 after a
change in temperature. At temperature $\tau = 2$, the tiles in
Figure \ref{figure:GadgetTiles} uniquely assemble a supertile
containing the 0 tile in the top right corner.  By raising the
temperature to $\tau = 5$, the two tiles in the east column that
lie north of the $z$ tile fall off and are replaced by two new
tiles, the 1 tile being placed in the top right corner where the 0
tile was originally.  Thus, the system uniquely assembles a
supertile with tile $0$ in the top right corner for temperature
sequence $\tau = \langle 2\rangle $ , and uniquely assembles a
supertile with tile 1 in the top right corner for temperature
sequence $\tau = \langle 2,5\rangle $.

In addition to permitting a single bit of information to be encoded
into an assembled supertile, bit flip gadgets can be used to control
the length of an assembled supertile.  That is, the default 0 tile
can constitute a \emph{dead-end} in that no more tiles can attach to
it, whereas the flipped 1 tile leads to the assembly of more tiles,
possibly another copy of the bit flip gadget.  The newly formed
gadget can in turn be set to a dead-end, or flipped to continue
growth. Combining these two uses for gadgets, we now show how a
slightly modified version of the basic bit-flip gadget permits the
assembly of arbitrary binary strings of arbitrary length.
\vspace{5ex}
\subsection{Building Arbitrary Binary Strings with $O(1)$ Tiles}
\label{subsection:babst}
\begin{figure}[!t]
\begin{center}
\includegraphics[scale=.45]{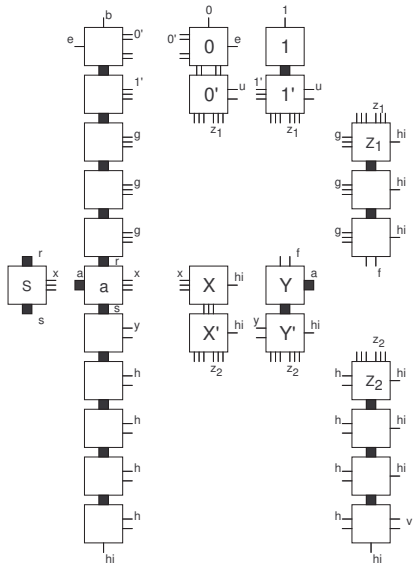}
\caption{These tiles implement a modified bit flip gadget that
constitutes a tile set of constant size that can uniquely assemble a
supertile that represents any arbitrary binary string of any length
$m$, where the string assembled is determined by a length $O(m)$
temperature sequence.  The dark black lines denote strength 9 glues.
The glues 0, 1, $b$, $u$, $v$, and $hi$ are not used to assemble the
binary string, but are used when the tile set is extended to
assemble $n\times n$ squares in Section \ref{section:bast}.}
\label{figure:StrongWeakGadget}
\end{center}
\end{figure}

\begin{figure*}[!t]
\begin{center}
\includegraphics[scale=.48]{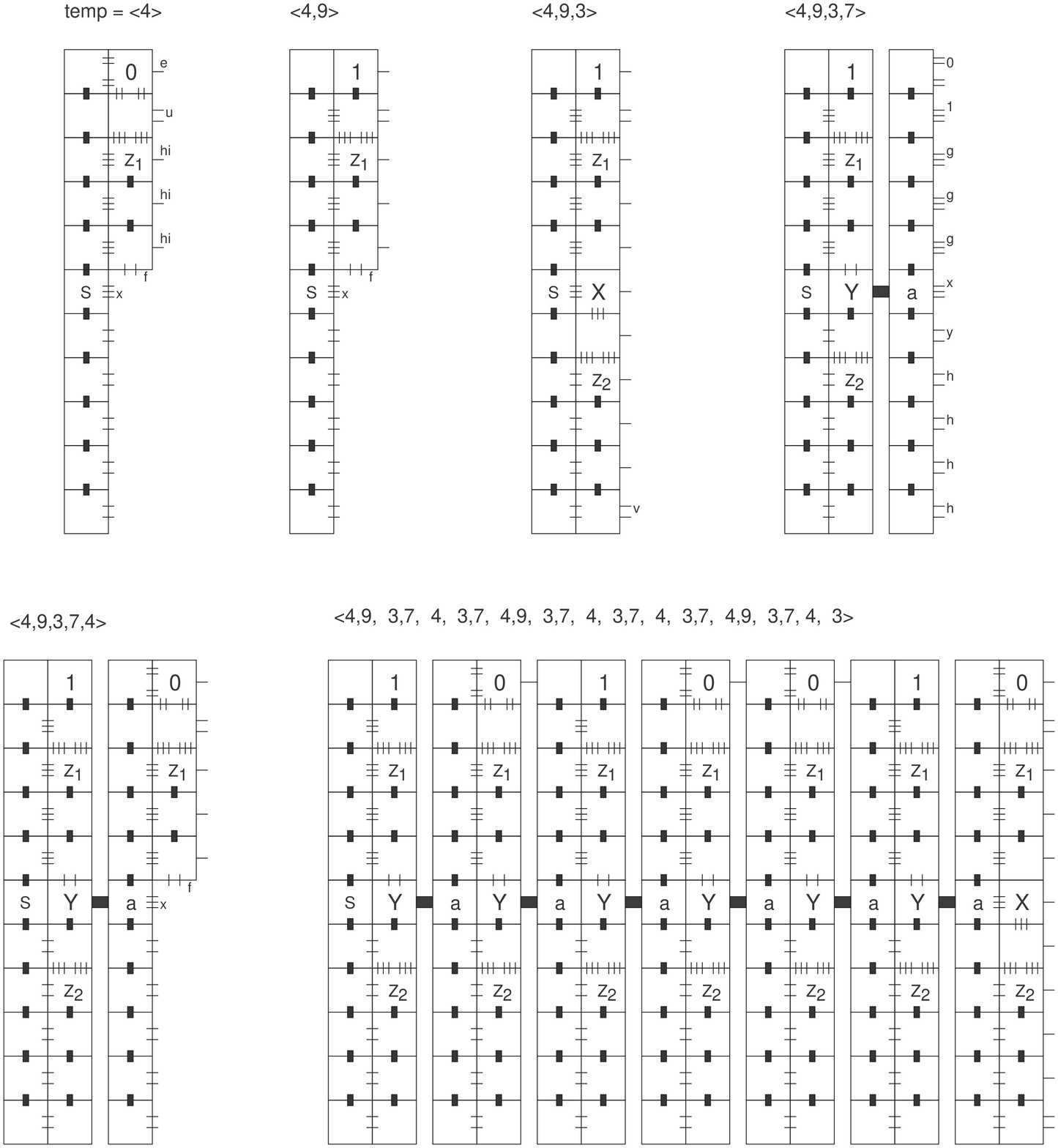}
\caption{A sequence of temperature changes to uniquely assemble a
supertile that encodes the string 1010010.}
\label{figure:BinaryExample}
\end{center}
\end{figure*}

To assemble a supertile which encodes an arbitrary binary string
$b_0 b_1 \ldots b_{m-1}$, we use a modified version of the bit flip
gadget given in Figure~\ref{figure:StrongWeakGadget}.  The top half
of this gadget encodes either a 0 or a 1, while the bottom half
encodes either an $x$ or a $y$.  See
Figure~\ref{figure:BinaryExample} for an example of how this tile
set assembles for a given temperature sequence.  The basic idea is
that the top half of the gadget will encode a single bit of the
given binary string. The bottom half, on the other hand, will be set
to the $x$ tile if the corresponding gadget is meant to encode the
final bit of the bit string, whereas the $y$ tile will serve the
purpose of creating another copy of the gadget to encode another
bit.  Thus, by flipping the bottom half of the gadgets, the length
of the bit string can be precisely controlled, while flipping the
top half of the gadget specifies the value of each corresponding
bit.

\begin{theorem}\label{theorem:binary}  For any given length $m$ binary string $b = b_0b_1 \cdots
b_{m-1}$, there exists a sequence of temperatures, all less than or
equal to 9, of length $O(m)$ such that the tile set of Figure
\ref{figure:StrongWeakGadget} uniquely assembles a supertile $S$
that encodes $b$.  Specifically,  $S$ is an $11 \times 2m$ rectangle
such that the tile in the top row, $2i + 1$ in from the west, is the
0 tile if $b_i = 0$ and the 1 tile if $b_i = 1$.
\end{theorem}

\begin{proof}
For any arbitrary length $m$ binary number $b = b_0b_1\ldots
b_{m-1}$, consider a corresponding length $4m-1$ temperature
sequence $\tau^b = \langle \tau_0,\ldots \tau_{4m-2}\rangle $
defined as follows.

$\tau_{4i} = 4$ for $i=0,\ldots, m-1$;

$\tau_{4i+1} = \left\{
\begin{array}{l@{\mbox{  if  }}l}4
& b_i=0;\\ 9 & b_i=1; \end{array}
   \right.$

$\tau_{4i+2}=3$ for $i=0,\ldots,m-1$;

$\tau_{4i+3}=7$ for $i=0,\ldots,m-2$.

As an example, for $b=010$, we have the corresponding temperature
sequence $\langle $4, 4, 3, 7, 4, 9, 3, 7, 4, 4, 3$\rangle$.
Conceptually, the temperature sequence is broken up into groups of
four, where the sequences $\langle 4\rangle $ and $\langle
4,9\rangle $ encode a 0 or a 1 respectively into the top half of the
gadget, and the sequence $\langle 3,7\rangle $ flips the bottom half
of the gadget to produce another copy of the gadget. So, at
temperature 4, an $11\times 2$ rectangle is uniquely assembled with
the bottom half of the second column missing, and the two top tiles
of the second column being the 0 and 0' tiles. The resultant
supertile is stable at temperature 8, but tile 0 can be removed at
temperature 9, which in turn leads to the removal of 0'.  The 1'
tile can then attach at temperature 9 north of $Z_1$, followed by
the 1 tile.

Whether or not the temperature is raised to 9, by lowering the
temperature to 3, the lower half of the second column of the gadget
is filled in with the $x$ tile placed east of the seed.  By raising
the temperature to 7, the $x$ tile is exchanged for the $y$ tile and
the first column of a second gadget is assembled, with the $a$ tile
in place of the seed tile.  Further, notice that this new supertile
has become stable at temperature 9, rather than 8, since if the
initial gadget was set to 0, the first column of the second gadget
stabilizes it with an additional strength 1 bond.  By dropping the
temperature to 4, and optionally raising it to 9, a 0 or a 1
respectively is encoded into the second gadget. A drop to
temperature 3 fills in the rest of the gadget, and a raise to
temperature 7 begins a new gadget and stabilizes the previous gadget
at temperature 9.  This process continues until the entire $11\times
m$ rectangle is assembled with the corresponding binary number
encoded in the top row.

\end{proof}

\section{Building Arbitrary $n\times n$ Squares in $O(1)$ Tiles}
\label{section:bast}
\begin{figure*}[!t]
\begin{center}
\includegraphics[scale=.48]{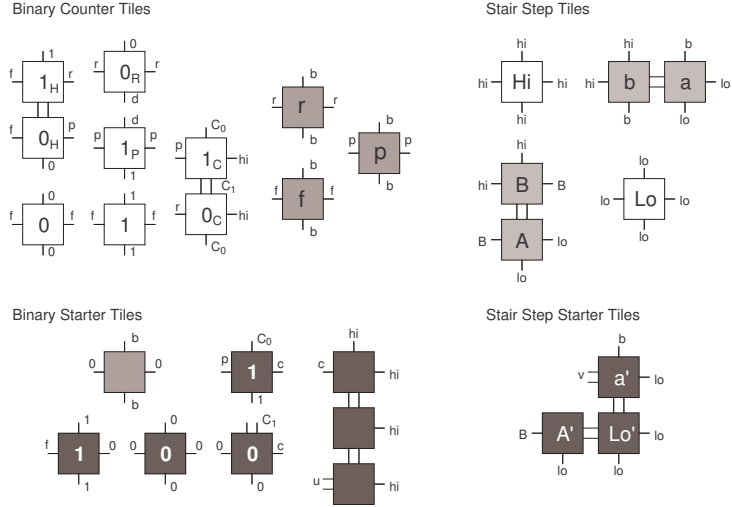}
\caption{A tile set for reading a binary string encoded into a
supertile and building a corresponding $n\times n$ square.}
\label{figure:SquareTiles}
\end{center}
\end{figure*}

\begin{figure*}[!t]
\begin{center}
\includegraphics[scale=.32]{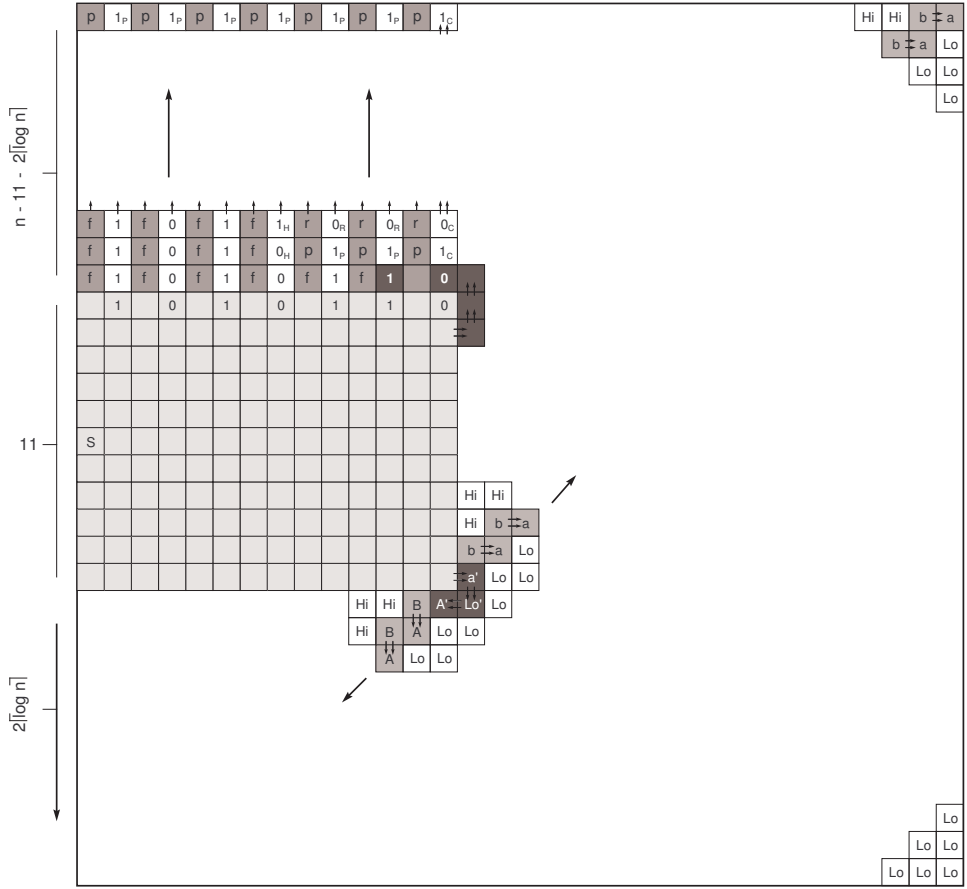}
\caption{The assembly of a $67\times 67$ square.}
\label{figure:SquareExample}
\end{center}
\end{figure*}

We now provide an application for the tile set of
Section~\ref{subsection:babst} for assembling an arbitrary binary
number in $O(1)$ tile complexity.  We combine this tile set with a
tile set for building $n\times n$ squares.  The basic idea is
similar to previous square building
constructions~\cite{Rothemund:2000:PSC,Adleman:2001:RTP} where a
constant number of square building tiles are combined with $O(\log
n)$ or $O(\frac{\log n}{\log\log n})$ tiles that assemble a $\log
n$ digit binary number.  The square building tiles then, in
essence, read the inputted binary number and assemble the square
uniquely described by the number.  These constructions require a
large tile complexity since they must build the input binary
string.  The tile complexity for assembly of almost all $n\times
n$ squares for a system that only changes temperature $O(1)$ times
is $\Omega(\frac{\log n}{\log\log
n})$~\cite{Rothemund:2000:PSC,Aggarwal:2004:CGM}. However, with
$O(\log n)$ temperature shifts we can build the binary number in
$O(1)$ tiles, and thus finish the entire square in $O(1)$ tiles as
well.

\begin{thmmain}
There exists a tile set $T$ of size $O(1)$ such that for any $n \geq
22$ there exists a temperature sequence $\{\tau_i\}_{i=1}^{k}$ of
length $k=O(\log n)$ such that $T$ uniquely assembles an $n\times n$
square with temperature sequence $\{\tau_i\}_{i=1}^{k}$.
\end{thmmain}
\begin{proof}
Consider the tiles from Figure~\ref{figure:StrongWeakGadget} with
the added tiles of Figure~\ref{figure:SquareTiles}.  For any $n\geq
22$, let $v= 2^{\lceil\log n \rceil}-n+11+2\lceil\log n\rceil$. Note
that $n=22$ is the smallest value such that $v<2^{\lceil\log
n\rceil}$.  Now, consider the length $4\lceil\log n\rceil-1$
temperature sequence of Theorem~\ref{theorem:binary} for encoding
the length $\lceil\log n\rceil$ binary representation of the integer
$v$.  To this sequence add a final drop in temperature to 2.  Thus,
the first $4\lceil\log n\rceil-1$ temperature changes uniquely
assemble an $11\times 2\lceil\log n\rceil$ rectangle encoding the
integer $v$.  At temperature 2, the binary starter tiles can attach
and begin the incrimination of a fixed length binary counter,
similar to the constructions in~\cite{Rothemund:2000:PSC}.  The
binary counter grows to a total length of $2^{\lceil\log n\rceil}-v
= n-11-\lceil\log n\rceil$.  With the addition of the length 11 seed
rectangle and the $2\lceil\log n\rceil$ positions filled in by the
stair step tiles, a width and length of exactly $n$ is achieved.
\end{proof}

From Theorem~\ref{theorem:square} and the fact that any square can
be assembled in $O(\frac{\log n}{\log\log n})$ tile complexity at
the fixed temperature 2 \cite{Adleman:2001:RTP}, we get the
following corollary.

\begin{corollary}\label{corollary:square}  For any $n\geq 1$, there exists a multiple
temperature tile system with tile complexity $O(1)$ and
temperature complexity $O(\log n)$ that uniquely assembles an
$n\times n$ square.
\end{corollary}

Thus, we have shown that all the tile complexity for the assembly of
an $n\times n$ square can be transferred into a length $O(\log n)$
temperature sequence.  A natural question to ask is whether this can
be reduced, or if there can be a smooth tradeoff between tile
complexity and temperature complexity.  The next theorem shows that
there is no tradeoff.

\begin{theorem}  For a tile set $T$ that uniquely assembles an
$n\times n$ square under temperature sequence
$\{\tau_i\}_{i=1}^{k}$, for almost all $n$ it cannot be the case
that both $|T| = o(\frac{\log n}{\log\log n})$ and $k=o(\log n)$.
\end{theorem}
\begin{proof}
The Kolmogorov complexity of an integer $N$ with respect to a
universal Turing machine $U$ is $K_U (N) = $ min$|p|$ s.t $U(p) =
b_N$ where $b_N$ is the binary representation of $N$.  A
straightforward application of the pigeonhole principle yields that
$K_U (N) \geq \lceil \log N \rceil - \Delta$ for at least $1-
(\frac{1}{2})^{\Delta}$ of all $N$ (see~\cite{Li:1997:IKC} for
results on Kolmogorov complexity).  Thus, for any $\epsilon > 0$,
$K_U (N) \geq (1- \epsilon)\log N = \Omega (\log N)$ for almost all
$N$.  This tells us that if we had a Turing machine that takes as
input a tile system and outputs the maximum length of the shape
produced by the given tile system, then the total size in bits of
the machine plus the size in bits of a tile system that uniquely
assembles an $N \times N$ square is $\Omega(\log N)$.  Such a
machine exists for the multiple temperature
model~\cite{Aggarwal:2005:CGM,Aggarwal:2004:CGM}, implying that an
encoding into bits of a tile system that uniquely assembles an
$n\times n$ square must be of length $\Omega(\log n)$ for almost all
$n$. \cite{Rothemund:2000:PSC} showed that a tileset and glue
function can be encoded in $o(\log n)$ bits if the number of tiles
in the set is $o(\frac{\log n}{\log\log n})$.  Therefore, since a
length $k$ temperature sequence can be encoded in $O(k)$ bits
(assuming, as we do, a constant temperature range), a tile system
with tile complexity $o(\frac{\log n}{\log\log n})$ and temperature
complexity $o(\log n)$ can be completely encoded into a length
$o(\log n)$ bit string, which cannot happen for almost all $n$.
\end{proof}

Thus, there are two possible versions of an optimal square building
assembly scheme.  The first is one which obtains $O(1)$ temperature
complexity and $O(\frac{\log n}{\log\log n})$ tile complexity, which
is achieved in \cite{Adleman:2001:RTP}, and the second is one which
obtains temperature complexity $O(\log n)$ and tile complexity
$O(1)$, which is obtained by Corollary~\ref{corollary:square}.

\section{Adding Robustness}\label{section:ar}
\begin{figure*}[!t]
\begin{center}
\includegraphics[scale=.50]{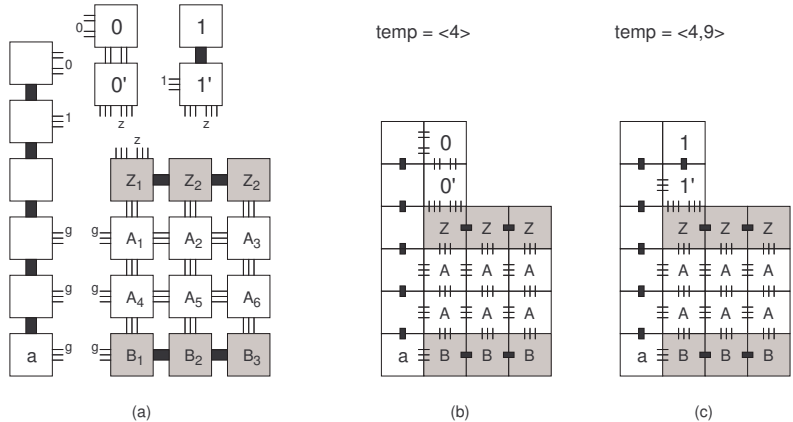}
\caption{(a)  Tiles that implement the multiple tile bit flip
gadget.  Dark black lines denote strength 9 glues. (b) At
temperature 4 a supertile with the 0 tile at the top of the second
column is assembled. (c) By raising the temperature to 9, the 0 tile
breaks off and is replaced by the 1 tile.}
\label{figure:GadgetMultiTile}
\end{center}
\end{figure*}

In this section we discuss how to modify the bit flip gadget
construction to make it robust against certain types of potential
assembly errors. In particular, the issue of concern is that the
standard tile assembly model assumes that tiles attach to the
growing seed supertile as singleton tiles.  However, it is plausible
to believe that in reality, singleton tiles could come together away
from the seed to form large supertiles, which could then
cooperatively attach to the seed to create a resultant supertile
that could not be created under the standard model. For example,
consider the bit flip gadget tiles in
Figure~\ref{figure:GadgetTiles}.  It is plausible to believe that in
implementation, the $z$ tile and the tile to its south could come
together independent of the seed tile and attach as a unit at only
temperature 2, even without the placement of the 0 tile, due to the
combined strength of the two strength 1 $g$ glues.
 If this were to happen, the construction would not be unique as
either a $0$ tile or a $1$ tile could then be placed in the top
right corner.  This non-uniqueness would prohibit the ability to
program a bit of information into the supertile.

To address this concern, we modify the gadget so that the assembly
is not only unique under the \emph{standard} tile assembly model,
but unique under a model in which tiles can come together and be
added as supertiles as well. Consider the following assembly model
introduced in~\cite{Aggarwal:2005:CGM,Aggarwal:2004:CGM}.
\\\\
\noindent \textbf{The Multiple Tile Model.}  In the \textit{multiple
tile} model tiles can combine into supertiles in a two-handed
fashion before being added to the growing seed supertile.  More
specifically, a tile set $T$ and a temperature $\tau$  have a
corresponding set of \emph{addable} supertiles $W(T,\tau)$.
Intuitively, this set constitutes the set of supertiles that can
potentially be attached to the growing seed supertile.  This set is
defined recursively.  First, $T \subseteq W$.  Second, for any two
supertiles $A,B \in W$, if $A$ and $B$ can be abutted together to
form a supertile $A \oplus B$ such that the total strength from the
glue function of all abutting edges of the two supertiles meets or
exceeds $\tau$, then $A\oplus B$ is also in $W(T,\tau)$.  Given this
set of addable supertiles, assembly takes place in the same fashion
as in the standard model, except that attachments to the growing
seed supertile come from the set $W(T,\tau)$, rather than just $T$.
For a more technical definition of the multiple tile model
see~\cite{Aggarwal:2005:CGM,Aggarwal:2004:CGM}.\\

From this definition, we see that the constructions so far in this
paper are not unique under this model.  However, we can make a
modification to the bit flip gadget construction so that its
assembly is unique.  The key is to replace the chain of tiles south
of the $z$ tile with a square block of tiles. Consider the tiles for
the \emph{multiple tile} bit flip gadget given in
Figure~\ref{figure:GadgetMultiTile}.  The first property of this
system is that the tiles that are placed south of the $z$ tiles are
placed cooperatively at temperature 4 with two weaker strength 3
glues. Thus, at temperature four, none of the tiles south of the $z$
tile can come together independent of the seed, and thus cannot be
placed before the placement of the initial 0 tile.

The second important property of this set is that there are three
columns of tiles south of the $Z$ tiles.  This serves the purpose of
holding the $Z$ tiles in place, as three distinct strength 3 glues
are needed to stabilize $Z$ at the higher temperature 9.  Finally,
note that the $Z_3$ and $B_3$ tiles are connected with strength 9
glues.  All tiles in the $4\times 3$ block of tiles including the
$Z$ row have at least 3 neighors with the exception of the two
corner tiles. These tiles thus need to be attached with strength at
least 6 on one of their two edges to maintain stability at
temperature 9.

By making modifications to the multiple tile bit flip gadget similar
in spirit to the modifications made to the original gadget, we get
corresponding constructions for $n\times n$ squares.

\begin{figure*}[!t]
\begin{center}
\includegraphics[scale=.50]{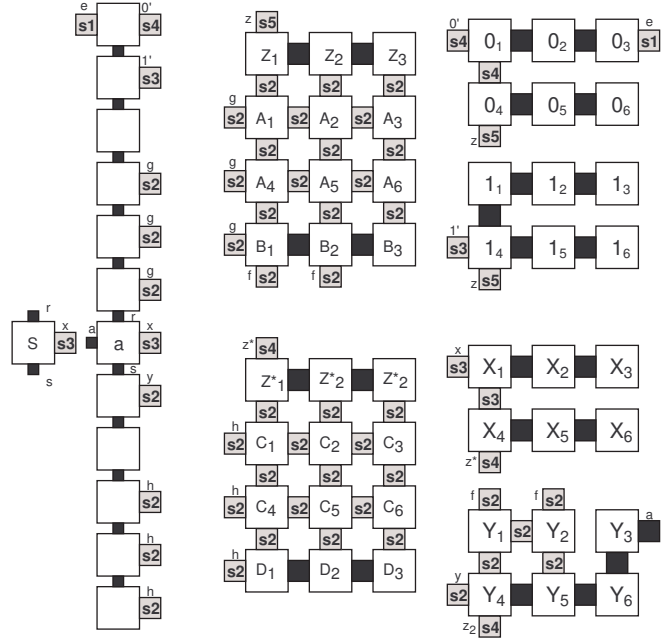}
\caption{These tiles constitute a tile set of constant size that can
uniquely assemble a supertile under the multiple tile model that
represents any arbitrary binary string.  The strength of each glue
is given by an index $s_i$.  Let $s_1=3$, $s_2=7$, $s_3=8$,
$s_4=10$, $s_5=13$, $s_6=17$ and $s_7=21$ where $s_7$ denotes the
strength of the glues denoted by dark black lines and $s_6$ is a
value used in the temperature sequence of the tile system.}
\label{figure:SWMultTile}
\end{center}
\end{figure*}

\begin{lemma}\label{lemma:squaremultitile}  Let $s_1,\ldots s_7$ be
positive integers such that:
\begin{enumerate}\small{
\item $s_1<s_2<s_3<s_4<s_5<s_6<s_7$;
\item $2s_4 < s_7\leq 2s_4 + s_1$;
\item $s_7\leq s_3 +s_5$;
\item $s_7\leq 3s_2$;
\item $2s_3 < s_6\leq s_2 +s_4$;
\item $s_4\leq 2s_2$.}
\end{enumerate}
Further, for any arbitrary length $m$ binary number $b =
b_0b_1\ldots b_{m-1}$, define the corresponding length $4m-1$
temperature sequence $\tau^b = \langle \tau_0,\ldots
\tau_{4m-2}\rangle $ as follows.

$\tau_{4i} = s_4$ for $i=0,\ldots, m-1$;

$\tau_{4i+1} = \left\{
\begin{array}{l@{\mbox{  if  }}l}s_4
& b_i=0;\\ s_7 & b_i=1; \end{array}
   \right.$

$\tau_{4i+2}=s_3$ for $i=0,\ldots,m-1$;

$\tau_{4i+3}=s_6$ for $i=0,\ldots,m-2$.

Then, the tileset $T$ given in Figure~\ref{figure:SWMultTile}
uniquely assembles a $12\times 4m$ rectangle under the multiple tile
model for temperature sequence $\tau^b$ such that the tile in the
top row, $4i+1$ in from the west, is labeled 0 if $b_i=0$ and 1 if
$b_i=1$.
\end{lemma}

\begin{proof}


As a base case, we first show that the first $12\times 4$ rectangle
is assembled correctly under the multiple tile model after the first
five temperatures changes in the temperature sequence $\tau^b$. Let
$\Upsilon_\tau$ denote the unique supertile assembled for some
temperature sequence $\tau$, assuming it exists.

For the initial temperature $s_4$, we can show that
$\Upsilon_{\langle s_4 \rangle}$ is a $12\times 4$ rectangle missing
the last three columns for the bottom six rows with the following
row by row composition: The last three columns of rows one and two
consist of 0 tiles, row three consists of $Z$ tiles, rows four and
five consist of $A$ tiles, and row six consists of $B$ tiles.  It is
straightforward that this is the case under the single tile model.
To show uniqueness under the multiple tile model, note that no
supertile in the addable tile set $W(T,s_4)$ contains two or more
rows, each row containing a tile with a glue that can bond to the
east edge of column one. Thus, since each west glue other than that
of a 1 tile has strength at most $s_3$, and $s_3 < s_4$, the first
supertile that can be attached to column one must correctly place
one or more of the 0 tiles. This then implies the unique assembly of
$\Upsilon_{\langle s_4\rangle}$.

After raising the temperature to $s_7$, the 0 tiles are removed
since $s_7$ is greater than $2s_4$ and $s_5$.  Further, nothing else
is removed since $3s_2 \geq s_7$ and any cut would require crossing
either three strength $s_2$ glues or one strength $s_7$ glue.  And
since $s_3 + s_5 \geq s_7$, the 1 tiles are placed where the 0 tiles
where originally.

After lowering the temperature to $s_3$, the bottom six rows are
uniquely filled in with $X$ tiles, $Z'$ tiles, $C$ tiles and $D$
tiles for reasons similar to those for the unique assembly of
$\Upsilon_{\langle s_4 \rangle}$.  After raising the temperature to
$s_6$, the $X$ tiles are removed since $s_6$ is greater than $2s_3$
and $s_4$. The tiles $Y_4$, $Y_5$, $Y_6$, and $Y_3$ are then placed
as $s_7\leq s_2 +s_4$.  Note that $Y_1$ and $Y_2$ do not yet have
enough strength to be attached.  These two tiles will be needed to
hold the bottom rows in place when the temperature is later raised
to $s_7$. However, they need to have glues fairly weak so that,
under the multiple tile model, they do not facilitate the placement
of the $Y$ tiles after the initial drop in temperature to $s_3$.
This weakness, as a side affect, does not allow their placement
until the temperature is dropped back down to $s_4$.

Finally, after the fifth temperature change back to $s_4$, $Y_1$ and
$Y_2$ attach to finish the first $12\times 4$ rectangle and permit
the replication of the original $\Upsilon_{\langle s_4\rangle}$
supertile just east of this rectangle.

Now consider the shape assembled after the first $4i$ temperature
changes of $\tau^b$ for some integer $i$.  Assuming the first $i$
dimension $12\times 4$ rectangles have been correctly assembled, the
$i+1$ rectangle will be assembled correctly after the first $4(i+1)$
temperature changes by the same analysis as for the first rectangle.
Thus, inductively the entire construction will be correct assuming
the first $i$ rectangles are stable at the highest temperature in
the remaining $(4m-1) - 4i$ temperatures of $\tau^b$. This is the
case as the highest temperature is $s_7$ and the glue $e$ east of
the first column of each rectangle stabilizes the previous $12\times
4$ rectangle at temperature $2s_4 +1$ in the case the 0 tiles remain
in place. Since $s_7\leq 2s_4+1$, the entire $12\times 4m$ rectangle
encoding $m$ is uniquely assembled for temperature sequence $\tau^b$
under the multiple tile model.

\end{proof}

\begin{theorem}\label{theorem:squaremultitile}  For any $n\geq 1$, there exists
 a multiple temperature tile system with tile complexity $O(1)$ and temperature
 complexity $O(\log n)$ that uniquely assembles an $n\times n$
 square under the multiple tile model.
\end{theorem}

\begin{proof}
Note that the integers $s_1=3$, $s_2=7$, $s_3=8$, $s_4=10$,
$s_5=13$, $s_6=17$ and $s_7=21$ satisfy the linear constraints in
Lemma~\ref{lemma:squaremultitile}.  The theorem thus follows from
Lemma~\ref{lemma:squaremultitile} and the observation that a trivial
modification to the tilesets in Figures~\ref{figure:SWMultTile} and
\ref{figure:SquareTiles} permit the extension from building an
arbitrary binary number to the building of an $n\times n$ square as
is done in Theorem~\ref{theorem:square}.  The uniqueness of assembly
under the multiple tile model for the binary counter tiles is shown
in~\cite{Aggarwal:2005:CGM}.
\end{proof}



\section{Future Work}
\label{section:future} There are many directions for future research
stemming from this work.  From an experimental perspective, it would
be interesting to implement and test the temperature programming
techniques of this paper with real self-assembly systems.  From a
theoretical perspective, we have shown that there exists a general
tile set capable of building any $n\times n$ square specified by a
sequence of temperature changes.  More generally, is it possible to
have a general shape building tile set that can be programmed to
assemble any arbitrary shape via a temperature sequence that encodes
a compact description of that shape? Another direction is to show
that there exists a tile set that simulates an arbitrary Turing
machine with its input specified by a temperature sequence. A third
direction is to consider how various error correcting techniques
used in other
work~\cite{Chen:2004:EFS,Chen:2004:ISA,Winfree:2004:PTS} can be
incorporated into temperature programmed assembly schemes. Another
open question is whether or not tile complexity and temperature
complexity can be reduced simultaneously by increasing temperature
range, which we have assumed to be bounded by a constant in this
paper.






\small
\bibliographystyle{abbrv}
\bibliography{all}

\end{document}